\documentclass[11pt]{article}
\usepackage{arxiv}
\usepackage{geometry}                
\usepackage{graphicx,bm}
\usepackage{epstopdf, epsfig}
\usepackage{xcolor,color}
\usepackage{amsmath}
\usepackage{mathtools}
\usepackage{amssymb}
\usepackage{amsthm}
\usepackage{hyperref}
\usepackage{cleveref}
\usepackage{siunitx}
\usepackage{longtable}
\usepackage[color=green!20]{todonotes}

\title{The mechanics and thermodynamics of tubule formation in biological membranes}
\author{Arijit Mahapatra*,
Can Uysalel*,
and Padmini Rangamani** \\
* both these authors contributed equally\\
** to whom correspondence must be addressed\\
Department of Mechanical and Aerospace Engineering, University of California San Diego,\\
9500 Gilman Drive, La Jolla, CA 92093, U.S.A.\\
Email address of corresponding author: \href{prangamani@ucsd.edu}{\color{blue}{prangamani@ucsd.edu}}}

\begin{document}

\maketitle

\begin{abstract}
Membrane tubulation is a ubiquitous process that occurs both at the plasma membrane and on the membranes of intracellular organelles. 
These tubulation events are known to be mediated by forces applied on the membrane either due to motor proteins, by polymerization of the cytoskeleton, or due to the interactions between membrane proteins binding onto the membrane. 
The numerous experimental observations of tube formation have been amply supported by mathematical modeling of the associated membrane mechanics and have provided insights into the force-displacement relationships of membrane tubes.
Recent advances in quantitative biophysical measurements of membrane-protein interactions and tubule formation have necessitated the need for advances in modeling that will account for the interplay of multiple aspects of physics that occur simultaneously. 
Here, we present a comprehensive review of experimental observations of tubule formation and provide context from the framework of continuum modeling. 
Finally, we explore the scope for future research in this area with an emphasis on iterative modeling and experimental measurements that will enable us to expand our mechanistic understanding of tubulation processes in cells. 

\vspace*{0.5cm} 

\noindent \textbf{Abbreviations} AC - anterograde carriers; BAR - Bin/Amphiphysin/Rvs; BDP - BAR domain protein; BFA - brefeldin A; BIN1 - Bridging Integrator 1; CICR - calcium-induced calcium release; ER - endoplasmic reticulum; ERES - ER exit site; ERGIC - ER-Golgi intermediate compartment; GFP - Green Fluorescent Protein; GUV - giant unilamellar vesicle; iPALM - interferometric photoactivated localization microscopy; LTCC - L-type Calcium Channel; PEC - Protrusion, Engorgement, and Consolidation; RyRs - Ryanodine receptors; SR - sarcoplasmic reticulum; TC - transport carrier; T-tubules - Transverse tubules; wtENTH - wild-type epsin N-terminal homology.
\end{abstract}

\section{Introduction}

The curvature generation capacity of biological membranes is critical for many cellular functions.
In the past few decades, the experimental studies of curvature generation in cellular and synthetic systems have given us physical insights into the underpinnings of curvature generation in membranes 
\cite{Bassereau2018}.  
Many of these studies have revealed the quantitative relationships between protein density, applied force, and the curvature generated \cite{Heinrich2010,Ford2002,Campas2009,Leduc2004}.

Curved membrane structures can broadly be classified into buds, pearled structures, and tubes \cite{Alimohamadi2018,yuan2020membrane}.
In this review, we focus on the formation of membrane tubes exclusively because of their broad application to membrane physiology. 
In eukaryotic cells, there are numerous applications of tubular protrusions at the plasma membrane. 
For example, a motile cell uses the actin-dense tubular structure, filopodia, to probe the environment during migration \cite{bornschlogl2013filopodial}. 
Filopodia also play a crucial role in neurite growth, the formation of dendritic spines, wound healing, and cellular trafficking \cite{mattila2008filopodia}. 
They  are also involved in cellularization in \textit{Drosophila} embryo \cite{sokac2008initiation} and adhesion of epithelial cells during embryo development \cite{vasioukhin2000epithelial}. 
Tubular protrusions from the plasma membrane also aid in the trafficking of cargoes (through transport carriers) \cite{polishchuk2003mechanism} and regulate trafficking of ions by restricting free diffusion with the help of their protective walls (in transverse tubules (t-tubules)) \cite{hong2014dmicrofl}. 
Beyond the plasma membrane, organelle membranes, such as the endoplasmic reticulum (ER) and the Golgi apparatus, can generate complex and dynamic tubular protrusions \cite{Lee1988,Mollenhauer1998,raote2020tango1}. 
Many of the molecular components involved in tube formation have been purified and vesicle-based systems have been used to study the underlying mechanisms.
The generality of tube formation in these processes, driven by protein crowding \cite{Stachowiak2010}, 
liquid-liquid phase separation \cite{feng2020}, 
osmotic pressure \cite{Sanborn2012}, polymer binding \cite{campelo2008polymer},
and even triblock copolymers \cite{Lim2017}, 
indicate that there are multiple ways to induce the compressive stresses associated with membrane tube formation.  

A critical aspect of research in the area of membrane mechanics is the close interactions between theoretical developments and experimental observations. 
For nearly five decades, the iterative development of theory, simulation, and experiment has resulted in a rich and vast literature spanning all areas of biophysics. 
In that spirit, we review some key highlights of tube formation in select experimental systems (Section 2), the associated mechanical models to explain these observations (Section 3), and the thermodynamic underpinnings of tube formation (Section 4).
We conclude with some critical open questions for future studies and suggest new interdisciplinary efforts in Section 5. 

\section{Experimental observations of membrane tubes}

In this section, we attempt to summarize the vast experimental literature into key mechanical aspects of tube formation in different biological conditions.

\subsection{Tubular protrusion in cells and their myriad functions}
\begin{figure}
\centering
\includegraphics[width=0.9\textwidth]{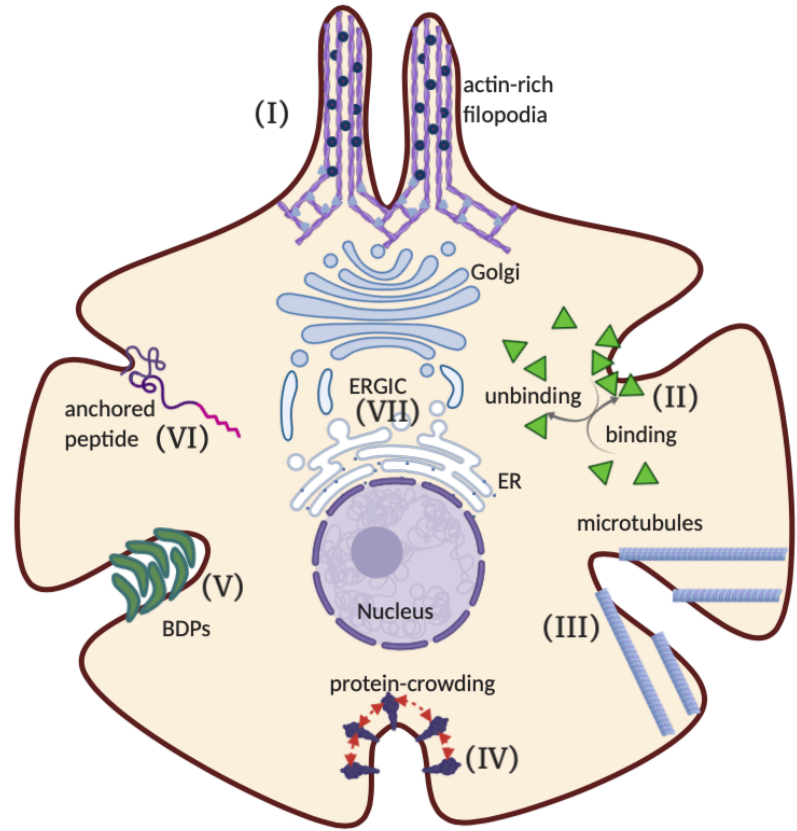}
\caption{Select mechanisms of membrane tubulation at the plasma membrane and in internal organelles. (I) Actin-driven filopodial protrusion, (II) tubular protrusion due to force generation caused by binding/unbinding of curvature-inducing proteins to the membrane, (III) tubular structure supported by microtubules in the cytoskeleton, (IV) tubular shape transformation of the membrane due to steric effect of crowded proteins, (V) spontaneous tubulation of membrane due to anisotropic intrinsic curvature induced by Bin/Amphiphysin/Rvs (BAR) domain proteins (BDPs), (VI) tubulation due to anchored motor protein or peptides, and (VII) tubular transport carrier (TC) during membrane trafficking in ER-Golgi intermediate compartment (ERGIC).
\label{fig:figure_1}}
\end{figure}

Tubular membranes are ubiquitously found at the plasma membrane and on intracellular organelles, and are implicated in a variety of cellular functions including membrane trafficking, cell migration, signaling, and probing the extracellular environment. 
These tubular structures are found in all eukaryotic cells. We present a few examples of these tubules to elaborate on their detailed structure and function relationships (Figure \ref{fig:figure_1}).

\subsubsection{Tubule formation at the plasma membrane}

\paragraph{\textbf{Filopodia:}}

Filopodia are finger-like cellular protrusions that play a crucial role in many cellular processes such as cell migration, axon and dendrite formation in neural growth, wound healing, and adhesion to the extracellular matrix. 
The structure of a filopodium is mainly supported by a bundle of actin filaments; the actin organization in the filopodium controls its length and elongation with the help of regulated polymerization and depolymerization of actin monomers \cite{theriot1991actin}.
Filaments from the lamellipodial actin network can elongate and come together at their barbed ends with the help of a tip complex (vasodilator-stimulated phosphoprotein (VASP)). 
Subsequently, proteins such as fascin assemble along the length and form a bundle of filaments that protrude a filopodia \cite{svitkina2003convergent,sekino2007precursors}.
The plasma membrane plays an important role in the formation of filopodia; the actin filament nucleating proteins (formins, Arp2/3, spire, etc.) bind to the membrane and induce the polymerization of actin filaments \cite{rohatgi1999interaction,co2007mechanism}, which results in tubular protrusions.
Additionally, there are instances of membrane deformation in filopodial-like precursors that can result in a nascent dendritic spine where a filopodial structure forms in the presence of membrane scaffolding protein (IRSp53, MIM, PSD-95) \cite{yamagishi2004novel,choi2005scaffold} and microtubules \cite{dent2007microtubules}. 
The interaction of the plasma membrane with a variety of regulating proteins that play an important role in the initiation and the regulation of the filopodial geometry is a common theme in different scenarios that induce the formation of filopodia.

\paragraph{\textbf{Tubule formation during membrane trafficking:}}
Eukaryotic cells have multiple internal organelles, each of which has a specific function.  
Proteins and lipids are transported from one compartment to another through membrane-bound organelles, called transport carriers (TC) \cite{stephens2001TC,wacker1997microtubule}. 
TCs can be made of small vesicles, single tubes, or complex tubular membrane structures \cite{nakata1998visualization,hirschberg1998kinetic}.  
In particular, tubule-shaped TCs can transport large cargo over longer distances when compared to vesicular TCs \cite{stephens2001TC}.
The mechanism of the formation of tubular TCs involves four basic steps \textemdash \hspace{0.05cm} budding of the membrane loaded with the cargo from the donor membrane, elongation of the tube, tubular fission, and finally, fusion to the acceptor membrane \cite{polishchuk2003mechanism}.
Membrane scaffolding proteins help the tubular protrusion at the donor site and subsequently support the elongation of the tube \cite{zhang2007clathrin}. 
Similar tubular elements are responsible for the transport of cargo through the endocytic pathway \cite{cullen2008endosomal}.

\subsubsection{Tubule formation in intracellular organelles}

\paragraph{\textbf{In Golgi-ER complexes ERGIC:}}
In mammalian cells, protein cargo is transported from the ER to the Golgi through a tubulovesicular cluster of membrane, which is often called the ER-Golgi intermediate compartment (ERGIC) \cite{hauri1992ergicbasic}.
This tubular structure is extremely dynamic in nature and works as a mobile transport complex that delivers cargoes from the ER to the Golgi \cite{horstmann2002dynamicERGIC}.
The complexity of transport in the ERGIC ranges from transport through a vesicle with a coat protein complex (COPI and COPII) \cite{aridor1995COPI} to the movement of large carriers along microtubules with the help of TCs 
\cite{wacker1997microtubule} and anterograde carriers (AC) \cite{ben2005AC} that contain fusion protein from ER exit site (ERES). 
Microtubules in the cytoskeleton interact with the tubular membrane and regulate these dynamics with the help of motor proteins in the early secretory pathway \cite{Koster2003}.
However, forces from the motor proteins alone are not enough to overcome the initial energy barrier of tubular protrusion \cite{koster2005force}; tubulation happens in the presence of GTPase  (IRSp53, CDC42 by activating ARP2/3 complex, Rac1) 
and other curvature generating agents \cite{bielli2005regulation}.
The ERGIC transport machinery also contains the SNARE-complex \cite{zhang2001ykt6}
and other tethering proteins \cite{gillingham2003tether} that help with transporting the multiprotein complex.  

\subsubsection{Select functions of tubules in whole cells}
We focus on some select functions of tubular structures in whole cells based on some of emerging research interests in cell mechanics. 
While not exhaustive, these functions give us some context on how the shape of the membrane tubule is closely tied to cellular function.
\paragraph{\textbf{Cardiac T-tubules:}}
T-tubules are tubular membrane structures that present in skeletal muscle cells and cardiac myocytes;
these tubular structures play a major role in muscle contraction. 
In cardiac myocytes, t-tubules invaginate from the sarcolemma and are organized along the z-discs surrounding the myofilaments \cite{hong2017cardiac}.  
T-tubules are organized in close proximity to the sarcoplasmic reticulum (SR) and assist in the rapid entry of Ca$^{2+}$ from the SR to the z-discs \cite{forssmann1970study,shepherd1998ionic}.
The L-type Calcium Channel (LTCC) on the membrane of the t-tubule stays in contact with Ryanodine receptors (RyRs) on the SR membrane \cite{scriven2000distribution,gez2012voltage} and is organized in dyad microdomains with the bridging integrator protein BIN-1 \cite{fu2016isoproterenol} 
that helps to stabilize the tubular structure \cite{hong2014dmicrofl}. 
This spatial organization of the calcium handling units in cardiac myocytes is thought to be important for the spatiotemporal dynamics of calcium in these cells \cite{forssmann1970study,soeller1999examination,nelson1963structural}. 

The tubular morphology of t-tubules is dynamic in nature and loss of tubules can occur in many disease states \cite{lyon2009loss} 
and can result in delayed kinetics of calcium-induced calcium release (CICR).
Even though the t-tubule structure dedifferentiates completely \textit{in vitro}, studies have confirmed that the tubular structure does not protrude as a result of forces applied to the membrane \cite{di2007mech}. 
Furthermore, these tubules are absent in stem cells prior to differentiation of cardiac myocytes \cite{lieu2009absence}, which suggests that the mechanism of t-tubule formation is yet to be completely understood.
A few studies suggest that the BDP Bridging Integrator 1 (BIN1) that attaches to the dyad \cite{hong2014dmicrofl,posey2014bin11,wu2010bin12,butler1997bin13}
is crucial to the formation of the tubular structure, indicating that membrane-associated curvature generating proteins play an important role in the formation of t-tubules.

\paragraph{\textbf{Neurons:}}
Another excitable cell type where the formation of tubules plays an important role is neurons.
Neuronal precursors undergo a series of morphological changes through tubular protrusions in multiple stages before they develop into a mature neuron \cite{kaech2006culturing}.
Early stages in these steps include the formation and elongation of smaller length scale filopodial and lamellipodial structures \cite{dotti1988neuritedev}. 
Many of the filopodial protrusions further elongate aided by actin-rich growth cone and form neurites \cite{dotti1988neuritedev}. 
In subsequent stages, one of the neurites undergoes further rapid elongation and develops into the axon, whereas the remaining neurites become dendrites.
The final stage consists of forming early dendritic spines (locations of synaptic contact) and axonal branches, which are protrusions at a smaller length scale.
Neuritogenesis, the process of neurite formation, is largely an actin-driven membrane deformation and the process happens in coordination with the actin cytoskeleton and membrane scaffolding proteins \cite{taylor2019opposing}.
The tubular geometries of neurites along with their electrical properties efficiently transmit the signals received from synaptic input to other cells \cite{miller1984neuronstr}.

Membrane tubule formation is also important at the small length scale for neuronal function.
Dendritic spines are small scale (length$\sim$1--5 $\SI{}{\micro\meter}$) 
protrusions along a dendrite that are sites of signal input from a neighboring neuron.
Similar to t-tubules, the tubular structure in spines is also very dynamic in nature and changes both with age and excitatory stage \cite{dailey1996dynfilop}.
The early spines are made of long and highly motile filopodial structures that seek a synaptic partner \cite{fiala1998synapparner}.
Eventually, the long filopodia develop into dendritic spines if the synaptic pathway strengthens and firings of neurons occur \cite{adams1982synapse}. 
These spines undergo structural changes with afferent input and in many cases disappear from the old location \cite{globus1966loss}.
This remodeling of spine morphology, known as structural plasticity, causes strengthening and weakening of synaptic contacts, which contributes to memory and learning \cite{barbosa2020memory}.

The growth cone, as mentioned earlier, is the actin-rich filopodial structure that elongates from the early filopodial structure to mature neurite and often produces a neural circuit in the brain \cite{ozel2015growthcone2}. 
The growth cone is very motile in nature and constitutes of three major structural regions \textemdash \hspace{0.05cm} an actin enriched peripheral domain often known as P-domain, a central domain consisting of organelles and microtubule, and a transition domain where actin interacts with microtubules \cite{ozel2015growthcone2}.
The entire structure flows and elongates at the same rate of axon elongation with the help of a Protrusion, Engorgement, and Consolidation (PEC) mechanism \cite{roossien2013growthcone1}. 
Thus, the plasma membrane plays pivotal roles in the structure and motility of the growth cone by assisting actin polymerization, receptor trafficking, recycling and turnover of membrane surface area, and adhesion to the extracellular environment \cite{tojima2011second}. 

\paragraph{\textbf{Development and Cellularization:}}   
Cellularization is the process that produces cell membranes for each nucleus in a \textit{Drosophila} embryo after they undergo mitotic division.
The plasma membrane of the embryo is covered with many finger-like small protrusions, known as microvilli \cite{Figard2013}. 
The formation of cleavage furrows is a critical step in development \cite{sokac2008initiation}. 
The cleavage furrows are thought to utilize the microvilli membrane reservoir to propagate alongside the nuclei and form compartments \cite{Figard2014}. 
These furrow canals contain proteins such as myosin 2, anillin, and F-actin, which are known to actively control the compartmentalization process  \cite{warn1983factin}.
Figard \textit{et al.} \cite{Figard2013} showed that the pulling forces of furrow ingression induce high plasma membrane tension; this tension can be sufficient to limit and/or stall actin polymerization at microvillar tips.
We note that microvilli unfolding depends on (a) interaction of the plasma membrane with BDPs, (b) interaction of the plasma membrane with actin filaments, and (c) membrane tension through regulation of furrow invagination and membrane trafficking.
The interaction between the plasma membrane, trafficking machinery, and force generating machinery is thought to be critical for the process of cellularization in \textit{Drosophila} embryogenesis. 
The use of force generating mechanisms to extend membrane tubules is commonly used to understand these force-displacement mechanisms.





\subsection{Tubule formation using forces and membrane-protein interactions} 


In this section, we focus on how the observations of tubule formation in cells has been studied in experiments with reconstituted systems to identify the biophysical mechanisms involved. 
Synthetic and reconstituted systems such as giant unilamellar vesicles (GUVs) are useful in studying the biophysical interactions of membranes and curvature-inducing components in a systematic manner. 
These systems also play a critical role in the iterative feedback between mathematical modeling and experimental observations \cite{sens2008biophysical,Sorre2009,Campas2009,Leduc2004}. 
A summary of the experimental observations and the underlying mechanisms is provided in Table \ref{tab:long}.



\subsubsection{Membrane forces and tubes} 
\label{sec:Membrane forces and tubes}

Forces exerted by motor proteins play an important role in membrane tubulation.
Several protein classes, such as motor proteins of the dynein and kinesin families, can mediate the interactions of membranes with microtubules (Figure \ref{Canfig2}a) \cite{Hirokawa1998}. 
For instance, kinesin motors bind to the membrane and pull tubular membrane protrusions while walking along the microtubules  \cite{Campas2009}.
According to \cite{Robertson2000,Lippincott1995}, \textit{in vivo} and \textit{in vitro} microtubule-based motor activity are both required in brefeldin A (BFA)-induced tubulation of Golgi membranes.
Fygenson \textit{et al.} \cite{Fygenson1997} observed changes in the shape of tubular protrusions in a vesicle that were caused by growth of a confined microtubule and showed the shape transformations in the buckling regime of microtubules. 
Further, motor proteins that bind to the membrane pull a tube after getting load support from the microtubules \cite{Roux2002}. 

There are many experimental measurements of pulling force and membrane parameters in tubular protrusion formation such as membrane tension and bending modulus (Figure \ref{Canfig2}b). 
Membrane tension is an important parameter that governs the force-displacement relationships of membrane tubes.
For example, Hochmuth \textit{et al.} \cite{Hochmuth1982} demonstrated that there is an inverse relation between the radius of tubular protrusion and the membrane tension. 
This kind of relationship can be verified mathematically by using the equilibrium formulation of membrane tube \cite{Derenyi2002}

\begin{equation} \label{equilibrium tube radius}
r =  \sqrt{\frac{\bar{\kappa}}{2 \sigma}}.   
\end{equation} 
Shao \textit{et al.} \cite{Shao1998} measured the critical pulling force in neutrophil and observed that when the force is below $34$ $\SI{}{\pico\N}$, the microvilli on the neutrophil membrane undergo small extensions.
However, when the pulling force exceeds $61$ $\SI{}{\pico\N}$, large tubular deformations occur.


Separately, the role of membrane tension and lipid flow was explored in substantial detail by a series of papers \cite{Koster2003,Leduc2004,Campas2009,Sanborn2012,Stachowiak2010,Simunovic2017,Rangamani2013}.
The dynamics of tube formation with a tether from cell membrane involves viscous drag caused by in-plane viscosity of the lipid, inter-monolayer friction, and friction offered by cytoskeleton. 
We find a series of experimental studies \cite{Hochmuth1982,Waugh1982,dai1995mechanical} and followed by theoretical analysis \cite{Hochmuth1996,Hochmuth1983} to find the viscosity of the membrane. 
Waugh \cite{Waugh1982} measured the viscosity of a phospholipid vesicle from a tether pulling experiment and observed the value of viscosity in the range of $5-13 \times 10^{-9}$ pN$\cdot$s/nm.
However, Hochmuth \textit{et al.} \cite{Hochmuth1982} measured viscosity of erythrocyte membrane from similar tether pulling experiments and reported the value of viscosity as $3 \times 10^{-6}$ pN$\cdot$s/nm.
Dai and Sheetz \cite{dai1995mechanical} observed the dynamic behavior of tube formation in a neuronal growth cone with a tether and pulled by optical tweezer.
They observed that the growth rate velocity of tether varies linearly with tether pulling force, which further confirms that the mechanics of tube formation is dominated by viscosity.
Hochmuth and colleagues \cite{Hochmuth1996} studied this force-velocity relationship of the growth cone tether analytically and reported that the effective viscosity is $1.37 \times 10^{-4}$ pN$\cdot$s/nm,
which contains three components \textemdash \hspace{0.05cm} in-plane viscosity, interbilayer slip, and cytoskeletal slip, with cytoskeletal slip making the most contribution. 

Further simplified systems, using GUVs alone, have been used to study force-displacement relationships. 
For example, tubular membrane protrusions can be induced from GUVs with the help of the external forces applied by optical tweezers \cite{Xu2008,Koster2003}. 
Koster and colleagues \cite{Koster2003} utilized optical tweezers to measure the forces that are involved in tubular protrusion formation.
The pulling force for tube formation measured in this study is significantly larger than the force applied by a single motor protein.
This led to the idea that multiple motor proteins assemble together to form a cluster that exerts enough force to extrude a tube. 
Going further, this study also elaborated on the role of membrane tension and showed that low tension is favorable for tube formation. 
An added layer of complexity arises due to the liquid nature of the bilayer; motor proteins can diffuse laterally on the vesicle.
Klopfenstein \textit{et al.} \cite{Klopfenstein2002} showed that certain kinesin motor proteins can bind to specific lipids directly and they can induce a dynamic preclustering mechanisms. 
These studies highlight how the dynamics of interaction between motor proteins and lipids plays an important role in the force generation mechanisms for tubule formation.

Leduc and colleagues \cite{Leduc2004} studied a biomimetic system which involved GUVs, kinesins, and microtubules. 
They presented both theoretical and experimental results that elucidated the dynamics of membrane tube formation, growth, and stalling. 
The results established that as kinesins can individually apply a pulling force of only 6 $\SI{}{\pico\N}$ \cite{Visscher1999}, molecular motors act collaboratively to induce tubes \cite{Campas2009}. 
These motors are able to pull membrane tubes and tube formation depends on both motor protein density and membrane tension.
Roux and colleagues \cite{Roux2002} demonstrated that typically between 15 and 30 motors are in contact with microtubules while inducing such tubular protrusions.


\subsubsection{Tubule formation from membrane-protein interactions} 


Next, we focus on observations in reconstituted systems for curvature generation by protein interaction with the bilayer. 
There are many proteins with specific domains that are known to induce membrane curvature \cite{Ford2002,Busch2015,Stahelin2003}.
Proteins such as endophilin \cite{Farsad2001} and amphiphysin \cite{Takei1999} bind directly to membranes through lipid binding domains. 
Such proteins can also generate tubular protrusions from liposomes \textit{in vitro} \cite{Farsad2001,Takei1995,Takei1998}. 
Protein-induced membrane bending generates the curvature of clathrin-coated pits and caveolae. During clathrin-mediated endocytosis, epsin family proteins can insert amphipathic helices in the cytoplasmic membrane leaflet \cite{Stachowiak2010}. 
It was also hypothesized that caveolins deform the bilayer through application of steric pressure \cite{Sens2004}. 
To explore the protein-lipid interactions in membrane protrusions, Stachowiak and colleagues generated a model system using GUVs and revealed that lipid domains can concentrate protein binding interactions, which can lead to the formation of tubular protrusions. 
Stachowiak \textit{et al.} \cite{Stachowiak2010} showed that tubular protrusion formation depends on the presence of fluid-phase lipids in the domain and requires a high density of protein attachment. 
These experiments led to a quantitative observation that tubule length has a linear relationship with vesicle diameter and a specific protein structure is not a requisite for tubular protrusion formation. 
Girard \textit{et al.} \cite{Girard2004} investigated the role of protein content in tubular protrusion formation during the reconstitution of membrane proteins into GUVs. 
Roux \textit{et al.} \cite{Roux2010} showed that dynamin-like proteins can deform membranes into tubular protrusions.
McMahon and colleagues \cite{Stahelin2003,Ford2002} showed that epsin, dynamin, amphiphysin, and endophilin can induce liposome tubulation independently \textit{in vitro}.
Leduc \textit{et al.} \cite{Leduc2004} conducted experiments on dynamics of motors and tube growth, and observed tubular protrusions \textit{in vitro} by kinesin motors that are in contact with GUVs and microtubules, establishing the role of membrane tension and motor density in tubular protrusion formation. 
These experiments established the ubiquity of tubule formation using different mechanisms.

\begin{figure}
\centering
\includegraphics[width=0.9\textwidth]{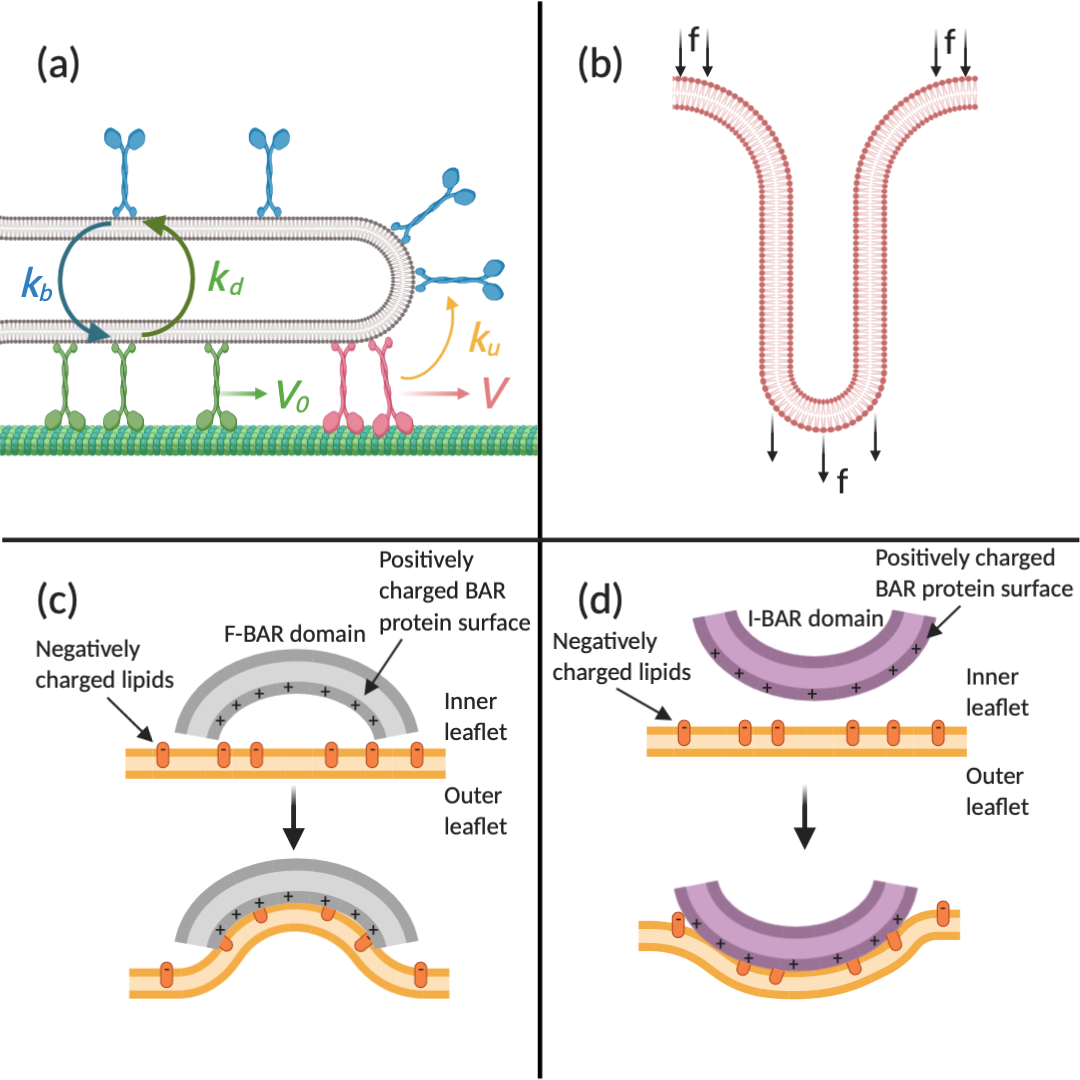}
\caption{\textbf{(a)} Schematic of a growing tubular protrusion (brown) along a microtubule (green). The motors are attached to the membrane and they can be either bound (green and pink) or unbound (blue) from the microtubule. When bound motors are far from the tip (green), they move with velocity $V_0$ and detach at a rate $k_d$. Unbound motors reattach to the tube at a rate $k_b$. The bound motors at the tip (pink) detach at a rate $k_u$. The tube growth velocity is $V$. \textbf{(b)} Tubular protrusion formation by forces that are exerted by cytoskeleton. \textbf{(c)} Illustration of binding mechanism of F-BAR domain protein (grey) to a lipid bilayer (yellow) that generates membrane invagination. \textbf{(d)} Illustration of binding mechanism of I-BAR domain protein (purple) to a lipid bilayer (yellow) that generates membrane exvagination.  \label{Canfig2}}
\end{figure}

Stachowiak and colleagues studied tubular protrusion formations with protein densities on membrane surfaces by exposing GUVs to wild-type epsin N-terminal homology (wtENTH) \cite{snead2017fission}.  
They showed that tubular protrusions are generated by the lateral pressure that is generated by collisions between bound proteins and steric congestion on cellular membranes \cite{Stachowiak2012}.
They also demonstrated how steric interactions between proteins can induce membrane bending \cite{Stachowiak2010}. 
Protein crowding on lipid domain surfaces forms a protein layer that buckles outward, this buckling bends the domain into stable tubules spontaneously. 
Lipid domains can confine protein binding on vesicle surfaces and protein binding can generate tubular protrusions by using two global parameters: domain size and membrane tension. 
Peter \textit{et al.} \cite{Peter2004} studied BDPs, which are anisotropic crescent-shaped proteins that have an intrinsic curvature.
BDPs form a banana-like dimer and the curved structure of these proteins provide them the ability to peripherally adhere to the membrane surface \cite{Frost2008}.
BDPs can bend the membrane in what is known as the scaffold mechanism. 
An additional mechanism that has been proposed for BDP-induced tubulation is the amphipathic wedge mechanism, which proposes that curvature is induced as a buckling response to the insertion of amphipathic sequences into the leaflet of the bilayer. 
The adhesion of the F-BAR domain protein to the lipid bilayer induces positive curvature (Figure \ref{Canfig2}c), while the adhesion of the I-BAR domain protein to the lipid bilayer induces negative curvature  (Figure \ref{Canfig2}d) \cite{Kabaso2011}.
These features can be captured by the curvature deviator model, which will be discussed in Section \ref{sec:model_BDP}.

\subsubsection{Role of membrane tension in tubule formation}
Here, we summarize some of the contradictory observations on the influence of membrane tension in tubule formation. In force-mediated tubule formation, higher membrane tension acts adversely to the length of the tube, and Der{\'e}nyi \textit{et al.} \cite{Derenyi2002} found that the tube length $L$ varies inversely with the membrane tension.
In GUVs, multiple studies have demonstrated that high tension requires higher force values to obtain tubes of a given length and radius \cite{Walani2015}. 
Changing the osmotic pressure is a classic method for changing the membrane tension. 
However, when osmotic pressure is present, an apparent contradictory nature of tube morphology with tension is observed.  
For example, Sanborn and colleagues \cite{Sanborn2012} found that a protruded tube in a GUV remains as a tube in negative osmotic gradient (corresponds to positive membrane tension) but takes pearling-like shape transformations in positive osmotic gradients (negative membrane tension).
How can we understand this behavior? In addition to tension due to osmotic pressure, in GUVs, surface-to-volume ratio is another physical parameter, which plays an important role. 
The GUV in their experiments already contained tubular extension from their surface.
When vesicles experience negative osmotic pressure, the volume enclosed by the vesicle is increased when compared to positive osmotic gradient. 
The tube-like shape which is already connected to the membrane shows pearling-like shape to enclose a lower volume for a given surface area.

The role of membrane folding and unfolding has been only explored to some extent in different experimental systems \cite{karatekin2003cascades} and in theoretical
models \cite{chabanon2017pulsatile}. 
A folded membrane corresponds to either very low membrane tension or negative membrane tension, which again has many consequences in force-deformation dynamics of the membrane.
For example, Steinkuhler \textit{et al.} \cite{steinkuhler2020super} showed that phase separation of the lipids softened the vesicular membrane and therefore undergoes deformation for a smaller force. 
Additionally, these vesicle contains large nanotubes in the lumen areas, and retraction of the tubular structures occurs after applying a tension to the membrane, which further supports the fact that positive tension can adversely affect the tubular morphology.
Furthermore, the surface area of the membrane can be altered by stretching, which releases the membrane tension locally.
In this case, notice that the experimental system not only has membrane stretching (altering the tension) but also membrane folding and unfolding are creating local reservoirs of surface area and possibly inhomogenous tension regimes.
Indeed, Shi \textit{et al.} \cite{shi2018cell}, recently in an experimental \textit{tour de force} showed that the membrane tension in cells is heterogeneous. 
Thus, this is a research topic that needs further investigation.

\section{Mechanics of tube formation}

The formation of tubular protrusions on membranes can be understood by considering the balance of forces on the membrane. 
We note that this mechanics approach is valuable for both equilibrium and dynamic configurations. 
The fundamental feature underlying many of these models is the elastic nature of the lipid bilayer.
The lipid bilayer is a thin elastic sheet, fluid in plane but solid in bending.
There have been significant advances in theoretical developments in the field of membrane mechanics \cite{Derenyi2002,Walani2014,AlimohamadiVasan2018,Iglic2013,Oster1982,Pearce2020,Rangamani2013,Simunovic2017,Leibler1986,Sens2006,Kabaso2012}.
We summarize them here with a specific focus on membrane tube formation. 


\subsection{Helfrich energy for membrane mechanics}
\label{sec:Helfrich}
The Canham-Helfrich energy \cite{Canham1970,Helfrich1973} is commonly utilized for modeling the elastic bending energy of lipid bilayers in membrane mechanics. 
This model proposes that the strain energy of a lipid bilayer can be written as a function of the surface curvatures.
The principle of virtual work tells us that the minimization of the strain energy will give us the equilibrium shapes of the membrane \cite{Helfrich1973,Jenkins1977,Steigmann1999}.
This model has been used to study many biophysical processes \cite{Julicher93,Sackmann86,Lipowsky91,Hassinger2017}. 
There are various mechanisms that govern the formation of curvature on the lipid bilayer in the protein-lipid interface. 
The proteins such as clathrin induce curvature due to wedging effect, whereas there are other proteins (ENTH) that induce an asymmetry between leaflets when they bind to lipid bilayers, which leads to a bending moment and results in curvature of the membrane.
This asymmetry between the leaflets is represented as a spontaneous curvature.
The strain energy per unit area is given by \cite{Helfrich1973}

\begin{equation} \label{classicalhelfrichperunitarea}
w =  \kappa(H - C)^{2} + \kappa_GK.  
\end{equation}



The total energy of the membrane is then given by

\begin{equation} \label{classicalhelfrichtotal}
W = \int\limits_{A} (\kappa(H - C)^{2} + \kappa_GK) dA, 
\end{equation} 
where $\kappa$ is the bending modulus, $H$ is the mean curvature of the membrane, which is the average of the two principal curvatures (Figure \ref{Canfig3}a), $C$ is the spontaneous curvature, $\kappa_G$ is the Gaussian modulus, $K$ is the Gaussian curvature, which is the product of the two principal curvatures (Figure \ref{Canfig3}a), and $A$ is the total membrane surface area \cite{Seifert1997}.

The sign of the Gaussian modulus governs the stability of the flat membrane.
Note that for a closed vesicle, the surface integral of Gaussian curvature remains constant as per the Gauss-Bonnet theorem \cite{Seifert1997}. 
Thus, the contribution of energy from Gaussian curvature is often neglected in the study of membrane bending \cite{Seifert1997}.

Fournier and Galatola \cite{Fournier97} showed that when the value of $\kappa_G$ was not in the range of $-2\kappa<\kappa_G<0$, the second order curvature elastic energy as we presented in Equation \ref{classicalhelfrichperunitarea} becomes negative and in that situation, fourth order components dominate.
The modified energy with fourth order terms in curvature leads to different shape instabilities, and a tubular shape is one of them.
However, for most lipid membranes, the values of $\kappa_G$ range from $-\kappa/3$ to $\kappa/2$ \cite{hu2012determining}.

In order to minimize the energy in Equation \ref{classicalhelfrichtotal}, constraints on the surface area are included.
Experimental observations of the membrane stretchability have revealed that the stretch modulus is quite high \cite{Kwok1981} and therefore, the membrane can be treated as effectively incompressible \cite{Rangamani2013interaction}. 
This incompressibility is imposed as a constraint \cite{Steigmann1999} and a Lagrange multiplier is used to mathematically represent this quantity \cite{Rangamani2020}, which is widely interpreted as membrane tension \cite{Steigmann1999}. 

\subsection{Tubule formation using forces and tension} 
\label{sec:Tubule formation using forces and tension}

A classic result using the Helfrich energy for membrane tube dimensions and how they are related to the applied forces was presented in \cite{Derenyi2002}.
We briefly summarize it here to demonstrate the utility of mechanical models in predicting quantitative relationships between the applied force and the tubule radius. 
Der{\'e}nyi \textit{et al.} \cite{Derenyi2002} studied membrane pulling with the point force $f$ and showed that the total membrane energy can be expressed as \cite{Derenyi2002} 

\begin{equation} \label{static}
E = \pi \bar{\kappa} \frac{L}{r} + 2\pi \sigma rL - fL,
\end{equation}
where $\sigma$ is the membrane tension, which is the Lagrange multiplier for area incompressibility as discussed in Section \ref{sec:Helfrich}, $r$ is the radius of tubular protrusion, $L$ is the length of tubular protrusion, and $\bar{\kappa}$ is the curvature modulus of membrane. 
Please note that the value of curvature modulus $\bar{\kappa}$ is $1/2$ of the value of bending modulus $\kappa$ for isotropic membrane, which was used in Equation \ref{classicalhelfrichperunitarea} and Equation \ref{classicalhelfrichtotal}. 
Minimizing the energy of a tubular protrusion with respect to $r$ and $L$ yields

\begin{equation} \label{staticpartial}
\\ \frac{\partial E}{\partial r} = -\pi \bar{\kappa} \frac{L}{r^{2}} + 2\pi \sigma L = 0,
\end{equation}

and

\begin{equation} \label{staticpartial2}
\frac{\partial E}{\partial L} = \pi \frac{\bar{\kappa}}{r} + 2\pi \sigma r - f = 0.
\end{equation}
Therefore, as we discussed earlier in Section \ref{sec:Membrane forces and tubes}, the equilibrium tube radius is given by  $\sqrt{\frac{\bar{\kappa}}{2 \sigma}}$ and the static force to hold the tube is $2\pi \sqrt{2\sigma \bar{\kappa}}$.
Please note that in these two expressions, we used membrane tension $\sigma$ as a free parameter. 
The curvature elastic framework of tube formation with a pulling force in an open membrane suggests that membrane tension remains constant in the domain and is obtained from its value at the boundary \cite{Derenyi2002,Steigmann1999}. 
However, the boundary tension is considered as lipid reservoir tension which often comes from self assembly energy of lipid molecules.

Powers \textit{et al.} \cite{powers2002fluid} performed a theoretical study followed by numerical simulation to predict the outcome of a classic experiment for a soap film in between two parallel rings with aligned centers, where the soap film is replaced by a lipid bilayer. 
The catenoid shape we see in the case of the soap film breaks down if the distance between the rings is high enough compared to the ring diameter. 
However, for an elastic lipid bilayer, the numerical simulation shows that for a sufficiently large distance between the rings, the lipid bilayer forms a long tubular connection. 
The tubular morphology remains there if one of the rings is taken with a lower diameter value compared to the other, which closely represents a tether on the membrane. 
The result shows that the tubular shape is one of the energy minima for elastic lipid-bilayer under a tether force or in the surface that connects two bodies as we find in the ER-Golgi connector (ERGIC).


\subsection{Modeling the interaction of membranes with BDPs in tubule formation}
\label{sec:model_BDP}

\begin{figure}
\centering
\includegraphics[width=1\textwidth]{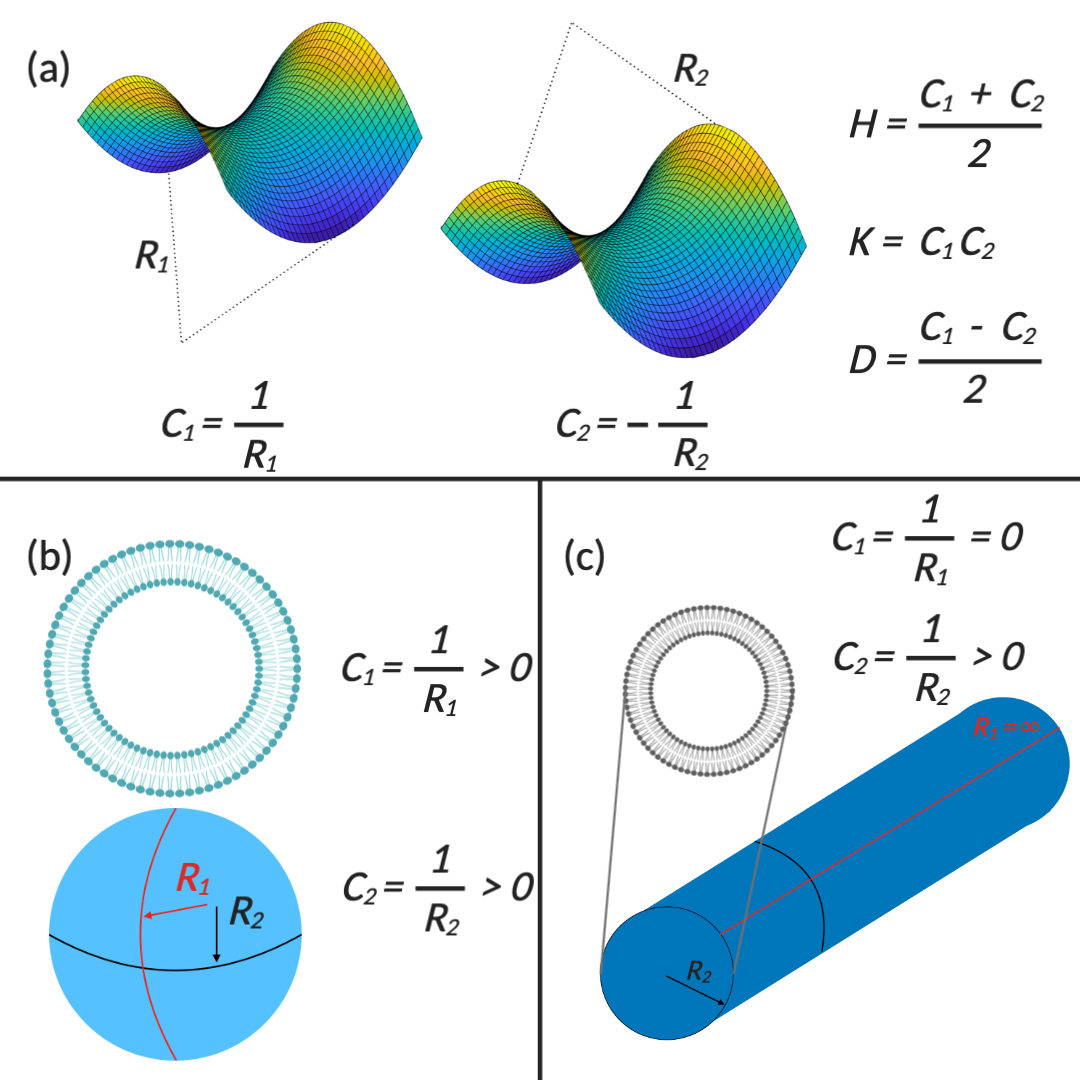}
\caption{\textbf{(a)} $R_1$ and $R_2$ are principle radii of a hyperbolic paraboloid surface and $C_1$ and $C_2$ are principal curvatures of a hyperbolic paraboloid surface. \textbf{(b)} Principal curvatures of a sphere. \textbf{(c)} Principal curvatures of a tube. \label{Canfig3}}
\end{figure}

Unlike a spherical shape (Figure \ref{Canfig3}b), to model a cylindrical tube formation, we note that for a cylindrical shape (Figure \ref{Canfig3}c), normal curvature along longitudinal axis is different from the normal curvature along the circumferential direction \cite{Frankel2011}.
Spontaneous curvatures generated by tubule forming proteins, such as BDPs, are inherently anisotropic in nature.

Therefore, the use of the isotropic spontaneous curvature model is insufficient for capturing the shapes of tubules and the relationship between tubule dimensions and protein densities on the membrane surface. 
To address this issue, a membrane strain energy density that captures the anisotropic curvature was proposed by many groups \cite{Walani2014,Alimohamadi2020,Fournier1996,Bobrovska2013}.
This modified Helfrich model was used for modeling the behavior of proteins that form tubular protrusions and induce an anisotropic curvature.
The energy per unit area in this case is written as \cite{Walani2014,Iglic2013,Iglic2005} 

\begin{equation} 
w = \quad \underbrace{\kappa_H(H - C)^{2}}_{\text{Elastic \quad  effects}} \quad \quad + \quad \quad \underbrace{\kappa_D(D - D_{0})^{2}}_{\text{Deviatoric \quad effects}}.
\end{equation} 
where $H$ is the mean curvature and $D$ captures the difference between the two principal curvatures (Figure \ref{Canfig3}a).
$D_0$ is the spontaneous deviatoric curvature,
$\kappa_H$ is the bending modulus for mean curvature, and $\kappa_D$ is same for the curvature deviator. 
For a linear elastic membrane, the value of $\kappa_H$ is same as the value of $\kappa_D$, which is $\kappa/2$ \cite{Walani2014}.
The values of $C$ and $D_0$ depend on the curvature and the orientation of an anisotropic protein which has different intrinsic curvature values in two principal directions.
For example, in Figure \ref{fig:figure_1}(V), BDPs orient along the circumferential direction of the cylinder.
In that case $C=D_0=r_0/2$, where $r_0$ is the intrinsic curvature of BDPs.
The total energy of the membrane is calculated as

\begin{equation} \label{modifiedhelfrichtotal}
\begin{split}
W = \int\limits_{A} \frac{\kappa}{2} \bigg[(H - C)^{2} 
+ (D - D_{0})^{2} \bigg] dA.
\end{split}
\end{equation} 
Note that for a linear elastic surface with no spontaneous deviatoric curvature ($D_0=0$), we recover the Canham-Helfrich expression of the energy as presented in Equation \ref{classicalhelfrichtotal}.

There are several applications that use the deviatoric curvature model that enhances our understanding of tube formation. 
Bobrovska \textit{et al.} \cite{Iglic2013} and Alimohamadi \textit{et al.}  \cite{AlimohamadiVasan2018} modeled tube formation by using the deviatoric curvature model to implement the effects of membrane elements and attached proteins with anisotropic properties. 
By using deviatoric curvature, Igli{$\check{c}$} and colleagues generated an anisotropy bending energy model for anisotropic membranes \cite{Iglic2005,Iglic2006,Iglic2002}. 
They used the deviatoric curvature model and observed that anisotropic membrane components play an important role in the stability of tubular protrusion formations.


\subsection{Current state of the art and future needs in dynamic measurements of tube formation in lipid membranes}

Thus far, we have focused on the equilibrium aspects of membrane tubule formation. 
We now turn our attention to the dynamic measurements of tubule formation.
Dynamic measurements of tube formation in lipid membranes can be achieved using optical tweezers; such optical tweezers are used to characterize the mechanical properties of the plasma membrane in terms of tether formation. 
According to Li and colleagues \cite{Li2002}, when compared to other tether formation techniques, optical tweezers provide noninvasive manipulation of cells with comparably great force resolution ($\sim$ 0.1 $\SI{}{\pico\N}$) and provide continuous monitoring of instantaneous tether force. 
Indeed, there is no dearth of data for dynamic measurements of tubule formation \cite{Oster1982,Pearce2020,Rangamani2013,Simunovic2017,Hochmuth1996,Sorre2009,Tian2009}.

There are also have been several models of the dynamics of tubule protrusion formation. 
Simunovic and colleagues modeled the dynamics of tube formation by mimicking the tubular protrusion formation. 
The corresponding experiments were done by pulling membrane nanotubes from GUVs using optical tweezers \cite{Simunovic2017}. 
Their model was based on balance laws and involved parameters such as externally applied force, tube area, change in tube area, tube length, change in tube length, and membrane tension. 
Simunovic \textit{et al.} \cite{Simunovic2017} combined their model with \textit{in vivo} and \textit{in vitro} experiments and demonstrated that motors provide tube pulling force and friction is an essential component for scission, which is the process of detachment of the protrusion from the plasma membrane.  


Subsequently, Hochmuth and colleagues \cite{Hochmuth1996} developed a thermodynamic analysis of the tether formation process and they developed experiments which were used to analyze neuron growth cones. 
They demonstrated that membrane viscosity is one of the important considerations for dynamics since it determines the rate of membrane deformation and it influences diffusion rates of particles in the surface plane \cite{Waugh1982}. 

Separately, based on experiments conducted in a multilamellar lipid system with osmotic pressure as a driver, Rangamani and colleagues developed a model including fluid drag, transmembrane pressure, and membrane tension along a tubular protrusion.
The model predicted that the three stages during tubular protrusion formation are initiation, elongation, and termination.
Based on experimental data, Rangamani \textit{et al.} \cite{Rangamani2013} constructed a mathematical model that can predict the tubular protrusion growth. 
They reported that their force balance approach can explain the elongation phase of tubular protrusion and that the confinement-based tubule growth system is regulated by osmotic pressure and drag. 
This simple force balance approach has also been used to explain the dynamics of elongation of acrosomes  \cite{Oster1982} and neurite retraction \cite{Pearce2020}.
The applications of this model to different processes have revealed that the membrane tension and the membrane viscosity are significant factors in governing the dynamic behavior of membranes. 
In certain cases, model predictions were verified experimentally.


\section{Thermodynamic considerations of tube formation}
Thus far, we have discussed the mechanical considerations of tube formation in lipid bilayers. 
The applied forces and membrane-protein interactions are also influenced by thermodynamic considerations and we briefly discuss them next. 

\subsection{Role of thermal fluctuations in tubule formation}
\label{sec:Thermal_fluctuation}
\begin{figure}[!h]
\centering
\includegraphics[width=0.8\textwidth]{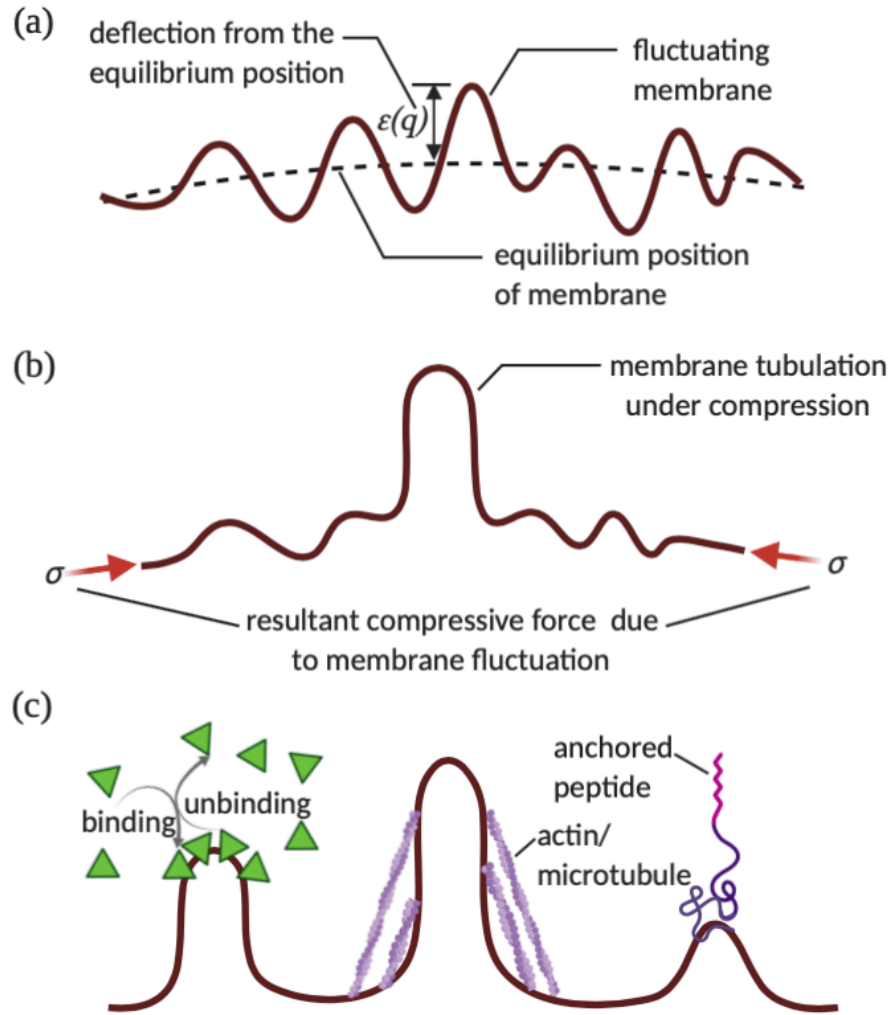}
\caption{\textbf{(a)} A fluctuating membrane with average mode amplitude deflection $\varepsilon(q)$ from the equilibrium position, \textbf{(b)} Membrane tubulation under compressive force caused by thermal fluctuation, \textbf{(c)} Tubulation due to active forces -- binding and unbinding of proteins, pulling of actin, microtubules and motor proteins on the cytoskeleton, and force induced by anchored and tethered proteins. 
\label{fig:figure4}}
\end{figure}

The bending energy of lipid bilayers is not high compared to the Boltzmann energy ($k_B T$) at physiological temperatures. 
As a result, lipid bilayers undergo shape undulations due to the thermal movement of the fluid molecules in the surrounding domain (Figure \ref{fig:figure4}a). 
Experimental observations have reported membrane fluctuations in vesicles 
\cite{patricia2005passive,fricke1984variation,brochard1975frequency}; 
these undulations cause mechanical softening of the membrane \cite{fricke1986flicker} and can influence shape instabilities in the bilayer \cite{shi2014fluctrev}. 
There are a series of theoretical studies \cite{helfrich1985effect,foster1986scale,peliti1985effects} and Monte-Carlo simulations \cite{gompper1996montecarlo} that have reported that thermal fluctuations soften the membrane a significant amount and also reduce local tension of the membrane.
The effective bending rigidity in the presence of thermal fluctuations can be written as \cite{foster1986scale}
\begin{equation}
\label{eqn:softening}
\kappa(T, \lambda, a) =\kappa_0-\frac{3}{4 \pi} k_{\mathrm{B}} T \ln \frac{q_{\mathrm{max}}}{q_{\mathrm{min}}},
\end{equation}
and the effective tension is given by \cite{foster1986scale,desernoinfluence}
\begin{equation}
\label{eqn:crumbling}
\sigma(T, a)  \simeq-\frac{3 k_B T}{8} (q_{\mathrm{max}}^2-q_{\mathrm{min}}^{2}),
\end{equation}
where $\kappa_0$ is the bending rigidity of the membrane in the absence of fluctuation, $q_{min}$ and $q_{max}$ are the magnitude of maximum and minimum wave numbers of the undulations respectively, $\lambda$ is the wavelength of the undulation, and $a$ is the diameter of lipid molecules. 
The equipartition of energy limits the energy of each undulation mode. 
Thus, the magnitude of the deflection correlates inversely with the square of the wavenumber of that particular mode of undulation.
Further, the ratio of these wavenumbers correlates with the maximum and minimum size of the wavelengths ($\lambda$) of the undulations as
\begin{equation}
    \frac{q_{\mathrm{max}}}{q_{\mathrm{min}}}=\frac{\lambda _{\mathrm{min}}}{\lambda _{\mathrm{max}}}.
\end{equation}
The highest value of the wavelength ($\lambda_{\mathrm{max}}$) is of the order of the length of the membrane ($L$), 
whereas the least value of it scales with the diameter of the lipid molecules ($a$). 

Considering Equation \ref{eqn:softening} and the fact that thermal softening is directly correlated with the size of the domain, the role of fluctuations can become prominent on a larger length scale. 
In contrast, for a lower length scale, the effect of thermal fluctuation will be negligible. 
The persistent length $\xi$ below which the membrane behaves as a rigid surface varies with \cite{peliti1985effects,de1982microemulsions}
\begin{equation}
\xi  \propto \left( \frac{4 \pi \kappa_{0} }{ 3 T}\right).
\end{equation}
The changes in physical properties of the membrane resulting from the effect of thermal fluctuations can facilitate shape instabilities, many of which lead to the formation of tubular protrusions \cite{shi2014fluctrev}. 
For low surface tension membranes, the shape undulation generates a negative tension and thus inserts a compression in the plane of the membrane.
As a result of this compression, the membrane undergoes a buckling instability resulting in the formation of a tubule out of the plane (Figure \ref{fig:figure4}b).
Such tubular structures have been observed in many experiments \cite{staykova2011comptubePNAS,solon2006comptube2}. 
Shape undulations also alter the binding probability of the molecules from the surrounding fluid \cite{marzban2017fluctbinding}, which confer additional surface area on the membrane and impose compressive stresses that support tubulation.  
The coupling between shape fluctuations and membrane-protein interactions can result in the clustering of proteins on the membrane surface due to in-plane attraction among the proteins \cite{pezeshkian2017mechanism}. 
These protein clusters can lead to tubulation of the membrane by means of a steric effect \cite{Stachowiak2010} or by spontaneous tubulation \cite{lipowsky2013spontaneous}.

\subsection{Thermodynamics of protein binding, aggregation, and phase separation}
The coupling between membrane mechanics and the thermodynamics of membrane protein interactions results in  thermophysical phenomena such as the aggregation of proteins,
the separation of protein and lipid phases, and
the binding and unbinding of proteins to the membrane. 
Proteins that do not interact with one another prefer a homogeneous distribution in the lipid bilayer to maximize the entropy of the system \cite{safran2018statistical}. 
However, proteins that interact with each other can experience a net attractive force among themselves and form a cluster \cite{schuster2018controllable}. 
Additionally, due to differences in chemical composition of the lipids, the protein-coated region can form a separate phase on the lipid bilayer \cite{heberle2011phase,lee2019lipid}.
The unbalanced force in the transition region induces line tension, which makes formation of a cluster of the same phase energetically favorable \cite{odell1981morpho,ursell2009morphology,garcia2007effect,baumgart2003imaging}. 

Theoretically, the effect of aggregation can be modeled by incorporating an aggregation potential in addition to the membrane bending energy \cite{leibler1987phase,weber2019aggrpot,gov2018guided}.
The binding and unbinding of proteins to the membrane and
adhesion of the proteins  
on the membrane can decrease the free energy of the system and are energetically favorable  \cite{reynolds2011thermobinding}.
Each of these thermophysical phenomena influences membrane bending and is conversely dependent on the membrane curvature created by bending (Figure \ref{fig:figure4}c). 
Veksler and Gov \cite{gov2007phase} presented a detailed theoretical model of filopodial protrusion where they considered the effect of protein adhesion, the force due to actin polymerization, and membrane tension separately on aggregation. 
They predicted that force due to polymerization increases the critical temperature of phase separation, whereas, adhesion strength and tension decrease the critical temperature of phase separation. 

Curvature plays a significant role in aggregation and phase segregation.  
We previously discussed in Section \ref{sec:model_BDP} how BDPs induce anisotropic curvature on the membrane and induce tubulation. 
Further, BDPs are flexible rod-like proteins that undergo elastic deformation in addition to inducing membrane curvature.
The energy for elastic bending of BDPs is thus dependent on the membrane curvature and minimizes the total combined energy of the membrane and BDPs with a preferable distribution of BDPs \cite{Iglic2006}.  
Another overlooked feature of these BDPs is that the anisotropic curvatures of proteins also induce an orientation entropy in the system.
A series of studies \cite{Iglic2013,Iglic2005,Iglic2006,Iglic2005ex,Perutkova2010} modeled the thermodynamics of BDP interaction with the membrane by considering the energy of bending of both the membrane and BDPs along with the entropy for configuration and orientation of BDPs. 
These studies suggest that BDPs undergo curvature-induced aggregation that eventually results in a tubular protrusion of the membrane. This tubular shape corresponds to the minimum energy of the system and the orientational entropy of BDPs favors this process \cite{Perutkova2010}. 

Another thermodynamic effect that influences tubulation of the membrane is protein crowding.
Protein crowding is the phenomenon that is associated with a high concentration of proteins in the lipid bilayers. 
When such macromolecules adsorb onto the membrane, they undergo steric interactions and impose an active tension to the membrane and often result in tubulation of membranes 
\cite{Stachowiak2010,stachowiak2010biomolecular}. 
Stachowiak \textit{et al.} \cite{Stachowiak2010} demonstrated this kind of tubular protrusion as membrane tension dominated, and proposed a physical model for the estimation of membrane tension ($\lambda$) as a function of protein-lipid binding energy ($\Delta G$)
\begin{equation}
    \lambda \approx \frac{3 \Delta G A_D}{A_P},
\end{equation}
where $A_D$ is the fractional area of the protein domain and $A_P$ is the fractional binding area of the protein.
They further assumed that membranes with crowded proteins undergo area dilation under this high tension.
Such crowded proteins can collide with each other and generate a lateral pressure which can lead to tubulation \cite{Stachowiak2012}. 
This crowding pressure, when sufficiently large, can also facilitate membrane fission \cite{snead2017fission,snead2019fission2}. 
Derganc and \v{C}opi\v{c} \cite{derganc2016membrane} theoretically modeled the curvature generation due to crowding pressure and estimated a spontaneous curvature ($C$) as a function of difference in crowding pressure between two monolayers, given by 
\begin{equation}
    C=\frac{h}{\kappa}\Delta p_c,
\end{equation} 
where $\Delta p_c$ is the difference in crowding pressure between two leaflets of lipid bilayer and $h$ is the distance between the neutral plane 
of lipid bilayer and the plane of steric repulsion, and $\kappa$ is the bending rigidity of lipid bilayer. 
Further, the crowding pressure is modeled in the same fashion as thermodynamic gas pressure, which encounters the effect of collision between the proteins.
This process of curvature generation with a crowding pressure is also considered as one of the basic thermodynamic mechanisms of curvature generation in protein-lipid interfaces \cite{stachowiak2013cost,Busch2015}.

\section{Future perspectives and open questions}
In the previous sections, we have elaborated on how cell-based experiments, model systems, and mechanical models have focused on the problem of membrane tubulation. 
Here, we discuss certain new avenues for this area of research and how we might be able to bridge some of the gaps between mechanics and cell biology. 

From a modeling standpoint, there is an increasing need for more sophisticated models that take multiple physical processes into account.
We are seeing an increase in extensions of models of membrane bending that go beyond the classical descriptors of spontaneous curvature, and include other features such as lipid viscosity and protein diffusion \cite{Rangamani2013,Mahapatra2020,Arroyo2019,Arroyo2009}. 
However, these models need to be brought closer to the experimental observations. 
A challenge that lies ahead is the development of numerical methods that are robust \cite{Vasan2020,sauer2017}. 
An additional opportunity lies in bridging molecular dynamics simulations to continuum mechanics simulations to build a truly multiscale model \cite{argudo2016continuum}. 

From an experimental standpoint, increasing the resolution of quantitative measurements in time and space in GUV based systems (\textit{e.g.} protein density, tubule radius, surface coverage) would provide invaluable data to constrain the free parameters in the model development process. Of course, as discussed earlier, dynamic measurements of the tubule formation process are critical for informing the relevant timescales in the models.

The next opportunity, in our view, lies in the gaps between models built for synthetic or purified systems and models for cellular processes.
For instance, Shtengel \textit{et al.} \cite{Shtengel2009} developed interferometric photoactivated localization microscopy (iPALM), a simultaneous multiphase interferometry that provides both molecular specification and resolution of cellular nanoarchitecture. 
Thus, there is an opportunity for the modeling community to interact more closely to work with large experimental data sets to identify the key physics underlying these processes. 
Finally, we would like to iterate that there are many opportunities that call for truly interdisciplinary collaborations with open science approaches that can help us gain more insight into the fundamental processes of tube formation in cellular membranes. 


\section*{Acknowledgments}
The authors would like to thank their many collaborators in the field of membrane mechanics for discussing ideas and the organizers of the International Symposium on Cell Surface Macromolecules 2020 for engaging discussions. 
They would also like to acknowledge Haleh Alimohamadi, Prof. Ali Behzadan, Miriam Bell, and Jennifer Fromm for providing their critical comments and feedback for the manuscript. 
This work was supported by NIH R01GM132106 to P.R.  



\centering
\begin{longtable}{ |p{3cm}|p{6cm}|p{5cm}|}
\caption{A synthesis of membrane tubulation observations, experiments, and corresponding theoretical analyses} \label{tab:long} \\

\hline \multicolumn{1}{|c|}{\textbf{Experimental system}} & \multicolumn{1}{c|}{\textbf{Observation}} & \multicolumn{1}{c|}{\textbf{Mechanism \& related theory}} \\ \hline 
\endfirsthead

\multicolumn{3}{c}%
{{\bfseries \tablename\ \thetable{} -- continued from previous page}} \\

\hline \multicolumn{1}{|c|}{\textbf{Experimental system}} & \multicolumn{1}{c|}{\textbf{Observation}} & \multicolumn{1}{c|}{\textbf{Mechanism \& related theory}} \\ \hline 
\endhead

\hline \multicolumn{3}{|r|}{{Continued on next page}} \\ \hline
\endfoot

\hline \hline
\endlastfoot

\hline
\multicolumn{3}{|c|}{\textbf{Force-mediated tubulation}} \\
\hline
GUV + kinesins + microtubules & Roux \textit{et al.} \cite{Roux2002} showed that motor proteins that bind to the membrane pull a tube after getting load support from the microtubules. Leduc and colleagues \cite{Leduc2004} found that these molecular motors are able to pull membrane tubes and tube formation depends on both motor protein density and membrane tension. & 
Motor proteins apply pulling force on the GUV while walking along the microtubules, which generates tubular protrusion \cite{Derenyi2002}.  \\
&  & \\
GUV + optical tweezer & Koster \textit{et al.} \cite{Koster2003} demonstrated that multiple motor proteins assemble together to form a cluster that exerts enough force to extrude a tube. & As kinesins can individually apply a pulling force of only 6 $\SI{}{\pico\N}$ \cite{Visscher1999}, molecular motors act collaboratively to induce tubes \cite{Campas2009}. Clustering of motor proteins on the membrane should be an important consideration in theoretical developments. \\
&  & \\
Neutrophil + tether & Shao \textit{et al.} \cite{Shao1998} showed that when the pulling force is below $34$ $\SI{}{\pico\N}$, the microvilli on the neutrophil membrane undergo small extension.
However, when the pulling force exceeds $61$ $\SI{}{\pico\N}$, a large tubular deformation occurs. & The force-elongation curve of the tubular protrusion contains a threshold limit below which tube length monotonically increase with pulling force. However, large tubular elongation occurs above that threshold limit and tube length increases at constant pulling force \cite{Derenyi2002}. \\
 &  & \\
Erythrocyte + tether & Hochmuth \textit{et al.} \cite{Hochmuth1982} revealed that membrane viscosity is one of the important considerations for dynamics of tubular protrusion formation. & Membrane viscosity is a significant factor in governing the dynamic behavior of membranes \cite{Rangamani2013}. Membrane viscosity determines the rate of membrane deformation and it influences diffusion rates of particles in the surface plane \cite{Waugh1982}. \\
 &  & \\
Neuronal growth cone + optical tweezer & Dai and Sheetz \cite{dai1995mechanical} showed that growth rate velocity of tether linearly varies with tether pulling force. Hochmuth and colleagues \cite{Hochmuth1996} studied this force-velocity relationship of the growth cone tether analytically and reported that the effective viscosity is $1.37 \times 10^{-4}$ pN$\cdot$s/nm,
which contains three components \textemdash \hspace{0.05cm} in-plane viscosity, interbilayer slip, and cytoskeletal slip, with cytoskeletal slip making the most contribution. & Mechanics of tube formation is dominated by the membrane viscosity \cite{Rangamani2013}. \\
\hline
\multicolumn{3}{|c|}{\textbf{Protein-mediated tubulation}} \\
\hline
GUV + histidine-tagged GFP & Stachowiak \textit{et al.} \cite{Stachowiak2010} demonstrated that the tubular protrusion formation depends on the presence of fluid-phase lipids in the domain and requires a high density of protein attachment. They also demonstrated how steric interactions between proteins can induce membrane bending \cite{Stachowiak2010}. & Large proteins experience steric repulsion when they are crowded in confined space and the resultant thermodynamic crowding pressure induces curvature on the membrane \cite{derganc2016membrane}. \\
 &  & \\
Liposome + endophilin & Farsad \textit{et al.} \cite{Farsad2001} showed that endophilin binds directly to membranes through lipid binding domains. Endophilin can also generate tubular protrusions from liposomes \textit{in vitro}. & Anisotropic membrane components can stabilize and induce the growth of the tubular protrusions \cite{Kabaso2012}. \\
 &  & \\
GUV + wtENTH  & Stachowiak \textit{et al.} \cite{Stachowiak2012} revealed that tubular protrusions are generated by the lateral pressure that is generated by collisions between bound proteins and steric congestion on cellular membranes \cite{Stachowiak2012}. & Protein crowding can induce tubular protrusions \cite{Bobrovska2013} and membrane curvature is stabilized in region of high protein density \cite{Leibler1986,Perutkova2010}. \\
\hline
\multicolumn{3}{|c|}{\textbf{Tension-mediated tubulation}} \\
\hline
GUV + sucrose solution to induce osmotic pressure gradient & Sanborn \textit{et al.} \cite{Sanborn2012} found that the negative osmotic gradient generates tension, which induces cylindrical protrusions and a protruded tube in a GUV remains as tube in negative osmotic gradient but takes pearling-like shape transformations in positive osmotic gradients. & Membrane tension is a regulator in dynamics of tubular protrusion formation \cite{Sens2006}. \\

\end{longtable}


\bibliographystyle{unsrt}  
\bibliography{sample}

\end{document}